# Noise and Spin-noise: Cable dependence of Optimal Tuning/Matching Conditions

Eli Bendet-Taicher[1], Norbert Müller[2], Alexej Jerschow[1]*

[1]Chemistry Department, New York University, New York, NY 10003, USA

[2]Institute of Organic Chemistry, Johannes Kepler University, Altenbergerstraße 69, 4040 Linz, Austria

* to whom correspondence should be addressed:
    Alexej Jerschow, New York University, Chemistry Department, New York, NY 10003, USA
    alexej.jerschow@nyu.edu

**Abstract**

Previous studies have shown that tuning/matching conditions optimized for transmission and detection can be significantly different for a variety of commercial NMR probes. In addition, it was also shown that by optimizing reception tuning (as opposed to typical transmission or reflection tuning) one may in some cases obtain sensitivity enhancements by as much as 25-50%. In earlier work, spin-noise and absorbed circuit noise signals have also been used to characterize reception optima. In this work, we show how the length of the coaxial transmission line cable between the pre-amplifier and the probe affects the positions of the reception tuning optimum, the radiation damping strength, induced frequency shifts, as well as, the shape of the spin-noise and absorbed circuit noise line shapes.

**1. Introduction**

Recently, measurements of spin-noise and absorbed circuit noise showed that there can exist a significant difference in the tuning conditions for optimal transmission vs. optimal reception on conventional NMR probes [1-4]. We refer to spin noise as the process of spontaneous emission of signals by the spin system, and absorbed circuit noise as the process of absorption of circuit noise by the spin system. The measurement of the interactions between the spins and circuit (Johnson) noise allowed one to use these internal signal sources as a measure for the behavior of the received signals [5]. This approach was in many instances easier to implement than measuring the reception tuning sensitivity directly, especially for samples containing large signals (such as those from bulk $H_2O$), where solvent suppression techniques are difficult to reproducibly implement over a range of tuning conditions. Examples of this tuning procedure have been shown in both ambient temperature and cryogenically cooled probes [1, 3], as well as with hyperpolarized spins [6, 7] and with solid-state NMR probes [3, 4]. In that work, the appearance of a symmetrical "dip" spin noise line shape was seen as the condition for optimal detection. In several cases, however, it was observed that such a noise line shape was impossible to attain. Under yet different conditions, it was found that the settings for obtaining a symmetric noise line shape did not always coincide with those for optimal reception.

In addition, tuning-dependent frequency shifts were observed in the measured spin noise spectra [2], and a link was sought to the earlier-described proton signal frequency shifts induced by radiation damping [8]. In this article, we draw connections between tuning, frequency shifts, radiation damping [9, 10], quality factors, and the observed noise line shapes. Most importantly, we discuss the influence of the cable length connecting the preamplifier and the probe circuit, and demonstrate the periodic effects in the parameters, in addition to the factors that influence optimal reception tuning, and tuning for symmetric spin noise line shapes.

## 2. Theory

**Noise Line shape.** The NMR noise line shape can be derived from the Nyquist treatment [11-13] as described by Sleator et al [14, 15], and Ernst and McCoy [16]. The total spin-noise power $W(\omega)$ can be expressed as

$$W(\omega) = q \frac{1 + a(\Delta\omega)\lambda_r^0}{[1 + a(\Delta\omega)\lambda_r]^2 + [d(\Delta\omega)\lambda_r + 2Q\Delta\omega_c/\omega_c]^2} \quad (1)$$

where $\Delta\omega$ is the resonance offset, $\Delta\omega_c = \omega_0 - \omega_c$ is the offset between the Larmor frequency $\omega_0$ and the nominal tank circuit resonance frequency $\omega_c = \frac{1}{\sqrt{LC_t}}$ (according to Figure 1), $q$ is a frequency independent factor depending on the circuit parameters and temperature, $Q$ is the circuit's quality factor, and

$$\lambda_r = 1/T_{rd} = \frac{1}{2}\eta Q \gamma \mu_0 M_Z \quad (2)$$

is the radiation damping rate [10], where $\lambda_r = \lambda_r^0$ when the spin and coil temperatures are the same (i.e. in an ambient temperature probe).

The absorptive $a(\Delta\omega)$ and dispersive $d(\Delta\omega)$ spectral components,

$$a(\Delta\omega) = \frac{1/T_2}{(1/T_2)^2 + (\Delta\omega)^2} \quad (3)$$

$$d(\Delta\omega) = \frac{\Delta\omega}{(1/T_2)^2 + (\Delta\omega)^2}, \quad (4)$$

define the line shape of the NMR noise power signal, which may yield either a "bump" signal or "dip" signal relative to the circuit thermal noise level or various mixed line shapes with the absorptive and dispersive contributions.

According to the treatment of Eqs (1)-(4), a symmetrical "dip" would be observed at the rf circuit's resonance frequency (i.e. when $\Delta\omega_c = 0$). In practice, it was found, however, that this line shape may be observed at considerable tuning offsets from the optimum determined by the conventional tuning procedure (conventional tuning optimum – CTO), where the reflection coefficient is minimized [2]. The tuning offset at which one observes the symmetric spin-noise or absorbed circuit noise line shape has previously been called spin-noise tuning optimum (SNTO), and it varies considerably between different preamplifier-probe combinations. Typical offsets between CTO and SNTO were found to range over hundreds of kHz. A detailed line shape analysis under a variety of conditions can be found elsewhere, including the case of cryogenically cooled probes [2] and hyperpolarized spins [7].

While the treatment leading to Eq. (1) is appealing for its simplicity and for ease of line shape analysis, it does not account well for the significant differences between the CTO and SNTO settings. This discrepancy can be traced to the fact that the derivation of Eq. (1) uses the approximation $|\Delta\omega_c| = \left|\omega_0 - \frac{1}{\sqrt{LC_t}}\right| \ll \omega_0$. In practice, this condition is often not satisfied. For example, for $Q = 400$, $L = 40$ nH, tuning to $\omega_0 = 2\pi 500$ MHz, and series matching to 50 $\Omega$, one obtains $|\Delta\omega_c| \approx 2\pi 20$ MHz. Using this condition, it would seem impossible to achieve a symmetric line shape unless the circuit is detuned by a frequency of that order of magnitude. In practice, a transmission line cable is also connected between the probe circuit and the preamplifier, which further transforms the noise voltage expressions such that the SNTO tuning offset can be altered, as shown below.

An analytical treatment of the full circuit as modeled in Figure 1 quickly becomes complicated, but a numerical analysis can be performed easily using the following steps:
1. Calculate the circuit impedance $Z_A$ at point A.
2. Consider the impedance transformation by the transmission line at point B.
3. Calculate the total noise voltage spectral density at the preamplifier input, including both the Nyquist noise contributions from the sample and the circuit, as well as the preamplifier noise sources.
4. Determine the voltage conversion at the preamplifier input of an emf induced in the circuit and calculate the signal-to-noise ratio.
5. The effect of radiation damping and induced frequency shifts is calculated by determining the current and relative phase that can flow through the sample coil. For this purpose, it is convenient to calculate the combined impedance of the circuit in series of the sample coil (including the preamplifier input impedance, and its transformation via the transmission line).

**Noise analysis.** The impedance at point A (after series matching) in the circuit of Figure 1 can be written as

$$Z_A = \frac{1}{iC_t\omega + \frac{1}{r_c + z_s + iL\omega}} - i\frac{1}{C_m\omega} \quad (5)$$

where the contribution to the impedance due to the nuclear spins can be written as

$$z_s = \frac{r_{s0}}{1 + i\Delta\omega T_2} \quad (6)$$

for a Lorentzian resonance line shape with amplitude $r_{s0}$ [11]. Classically, this term can be considered to arise from the Brownian motion of the rotating frame magnetization [11]. Equivalently, we could express the spin contribution via the real and imaginary susceptibilities as resistive and inductive elements in the circuit [14]. Note that the term $\Delta\omega$ here is the offset frequency from the Larmor frequency of the resonance. The frequency-independent (or broad-band) contributions of the sample to the circuit resistance and inductance can be lumped into $r_c$ and $L$ for simplicity.

Taking the real part of Eq. (5) and using the Nyquist expression for the noise, one obtains the equivalent of Eq. (1) if $|\Delta\omega_c| = \left|\omega_0 - \frac{1}{\sqrt{LC_t}}\right| << \omega_0$ is assumed.

Next, the transmission line is taken into account. The frequency-dependent impedance seen at the preamplifier (point B in Figure 1) is then [17]

$$Z_B = Z_0 \frac{Z_A + Z_0 \tanh(\gamma x)}{Z_0 + Z_A \tanh(\gamma x)} \tag{7}$$

via the standard impedance transformation expression for a lossy transmission line of length $x$ (in units of wavelength). A characteristic impedance of $Z_0$ (50 Ω in our case), and the loss factor $\gamma = \alpha + i2\pi$ were used, where $\alpha$ is the loss factor per unit wavelength (for the RG223 cable used here, a typical value is $\alpha = 0.0497/\lambda$ at $\omega = 2\pi\, 500$ MHz). At the small multiples of the wavelength employed here, however, the assumption of zero-loss cables would lead to a negligible error.

At this point, the circuit can be modeled as a combined noise source for the circuit and spin noise given by the voltage spectral density

$$N_{rs}^2(\Delta\omega) = 4kT\,\text{Re}[Z_B] \tag{8}$$

with $Z_B$ and $Z_p$ (the preamplifier input impedance) appearing in series.

For the signal-to-noise treatment we follow references [18, 19] with slight modifications allowing for complex preamplifier impedance and complex $Z_B$. The preamplifier may be modeled [20] as adding a random current noise $I_n$ and random voltage noise $V_n$ to the circuit, in addition to a noise-less input impedance $Z_p$. The total voltage at the preamplifier input (across $Z_p$) can then be obtained via

$$V_{Noise}^2 = \left(N_{rs}^2(\Delta\omega) + |I_n|^2 \cdot Z_B^2 + |V_n|^2\right)\left|\frac{Z_p}{Z_B + Z_p}\right|^2. \tag{9}$$

This expression is equivalent to Eq. [A19] of Ref [18], except that it also allows for a complex transformed circuit impedance $Z_B$ and a complex preamplifier impedance $Z_p$. It will be shown below that optimal signal-to-noise ratio (SNR) is achieved when $Z_B$ is real.

The spectrum of the combined noise from the coil resistance, the nuclear spins and the pre-amplifier can be simulated based on Eq. (9) and plotted vs. $\Delta\omega$. The preamplifier noise sources, as modeled here, add frequency-independent noise. For line shape analysis, one can neglect these contributions (but not for signal-to-noise ratio calculations). Eq. (9) alone, predicts a shifting of the SNTO position in a periodic fashion as a function of the line length $x$ as will be shown below.

**Signal-to-Noise analysis.** Following a pulse with flip angle $\theta$, let the amplitude of the emf induced in the receiving coil by the precessing nuclear magnetization be ξ. This voltage will be transformed to $\xi\sqrt{\frac{\text{Re}[Z_B]}{\text{Re}[r_c + z_s \cos\theta]}}$ at the input of the preamplifier (as calculated via a power balance). Typically, one is interested in evaluating the SNR for small signals away from the signals of bulk solvent, and $z_s$ becomes negligible compared to $r_c$. For simplicity, we therefore drop the $z_s$ term, and the voltage at the preamplifier input from this induced signal is

$$S = \xi\sqrt{\frac{\text{Re}[Z_B]}{r_c}} \cdot \left(\frac{Z_p}{Z_B + Z_p}\right). \qquad (10)$$

The SNR is then from Eqs. (8-10) given as the square root of the power ratios,

$$SNR = \frac{\xi}{\sqrt{r_c}}\sqrt{\frac{\text{Re}[Z_B]}{4kT\,\text{Re}[Z_B] + |Z_B|^2|I_n|^2 + |V_n|^2}}. \qquad (11)$$

It is remarkable at this point that the preamplifier input impedance, $Z_p$ drops out of the SNR calculation. Eq. (11) shows that SNR is always larger when Im[$Z_B$]=0, and differentiating with respect to Re[$Z_B$] and solving for Re[$Z_B$] gives the optimum impedance at point B as [19, 20]

$$Z_{B,opt} = R_{opt} = \left|\frac{V_n}{I_n}\right|. \qquad (12)$$

If the circuit (including the transmission line) is matched to this optimum impedance, the best available SNR becomes

$$SNR_{opt} = \frac{\xi}{\sqrt{r_c}\sqrt{4kT + 2I_n V_n}}. \qquad (13)$$

It is readily seen that the condition for optimal SNR, Eq. (12), depends on the noise properties of the preamplifier rather than optimal matching (to 50 $\Omega$).

Changing the transmission line cable length can perform such an impedance transformation according to Eq. (7), so that optimal SNR may be achieved. It is also seen that the optimal SNR is achieved when $Z_B$ is real-valued. A correlation between the preamplifier noise sources $I_n$ and $V_n$ can also be considered, but has little effect on $Z_{opt}$ [19]. The presence of the preamplifier degrades the SNR by a factor of $\sqrt{1 + \frac{I_n V_n}{2kT}}$ (preamplifier noise factor) [18]

**Radiation Damping and Resonance Frequency Shifts.** The noise and pulsed signals further show marked frequency shifts, arising from an effect known as frequency pulling, which can be explained by radiation-damping induced frequency shifts [21]. Others have reported that such frequency shifts depended on mistuning [8], and it seemed natural to draw this connection here. Following [21], one can trace the appearance of resonance frequency shifts to non-canceled reactive impedances in the circuit. To develop this viewpoint here, it is convenient to consider the preamplifier impedance as transformed via the transmission line at point $A$,

$$Z'_p = Z_0 \frac{Z_p + Z_0 \tanh(\gamma x)}{Z_0 + Z_p \tanh(\gamma x)}. \qquad (14)$$

This impedance appears in series with $Z_A$, hence the total series impedance of the circuit becomes

$$Z'_A = Z_A + Z'_p \qquad (15)$$

The phase angle **Error! Objects cannot be created from editing field codes.** of this impedance as defined via

$$Z'_A = |Z'_A| \exp(i\psi) \qquad (16)$$

depends of course on the tuning condition, the preamplifier impedance, and the transmission line length, and affects the relative phase of the current passing through the coil. Both amplitude and phase have a bearing on radiation damping, and it can be shown that the radiation damping time constant in this case is [21]

$$\tau = \frac{|Z'_A| Q}{\lambda_r \omega_0 L \cos\psi}. \qquad (17)$$

As a result of the nonzero phase $\Psi$, the current in the coil also leads to a resonance frequency shift of

$$\delta\omega_s = \frac{1}{|Z'_A| Q} \sin(\psi) \omega_0 \lambda_r L. \qquad (18)$$

The constant $\lambda_r$ depends on $Q$ as defined above. It should be pointed out, however, that this constant would represent $Q = \frac{\omega_0 L}{r_c}$, which we will distinguish from $Q'$, the measured $Q$ factor in the assembly with the transmission line, which will be shown to vary with cable length.

For an idealized circuit where $\omega_c = \sqrt{\frac{1}{LC}}$, and $\Psi = 0$ at $\omega = \omega_c$, one may further simplify the expressions and obtain the frequency shift in Hz as [8]

$$f_s \approx \left( \frac{\Delta v_0 \alpha Q}{1 + 4\alpha^2 Q^2} \right). \qquad (19)$$

Where: $\alpha = \frac{f_0 - f_c}{f_0}$. The term $\Delta v_0$ is the resonance line width at half height when $f_0 = f_c$, $f_c = \omega_c / 2\pi$, and $f_0$ is the resonance frequency of the protons when $|\alpha| \ll 1$.

For the realistic case including a transmission line and unmatched preamplifier impedance, however, one needs to consider Eqs. (14)-(18), where it is seen that a change in cable length will produce changes in the reactance of $Z'_A$, and thereby shifts in the resonance frequencies. Experiments showing these effects will be shown below.

## 3. Materials and Methods

All experiments were recorded on a Bruker Avance 500 MHz spectrometer (11.7 T) equipped with a 5mm high resolution triple resonance (TXI, H, C, N) ambient temperature probe (sample temperature 298.3 K) using a $H_2O/D_2O$ (9/1) sample.

NMR noise experiments were performed while the rf-pulse amplifier input cable was disconnected from the $^1$H-preamplifier in order to minimize the impact of electronic noise generated by the pulse amplifier and other spectrometer hardware.

Spin noise data were collected using a pseudo 2D acquisition sequence, acquiring one block of noise per row at a spectral width of 10 ppm. A total of 512 blocks were collected in this way. Each block was Fourier transformed individually to a complex-valued (phase sensitive) spectrum, which was converted to a power spectrum (accumulating the phase sensitive data would lead to a cancellation of the noise signal) and then the rows were summed up to produce a one-dimensional noise spectrum.

For the measurements of signal shifts and sensitivity, the pulse durations were optimized so that delivered rf power and flip angles remain the same for all experiments. The pulse durations ranged from 7 μs to 21 μs for $90^0$ pulses and 0.4 μs to 0.6 μs for $5^0$.

For sensitivity measurements, a single pulse sequence with presaturation for water suppression was performed on a 2mM sucrose in a $H_2O/D_2O$ (9/1) sample. The signal to noise ratio was measured on the anomeric doublet left of the water signal over a noise range of 1.5 ppm between 5.5 ppm and 9.5 ppm.

All coaxial cables were RG223/U, 50 Ω, manufactured by Pasternack Enterprises, Inc. The connectors were female/male 50 Ω RF coax cable connectors, manufactured by Amphenol. The connectors were crimped to the coax cables cut to the desired lengths. Different cable lengths were made in increments of 0.1λ of $2\lambda \leq l \leq 3\lambda$, where the calculated wavelength was λ=39.54 cm. The cable length was measured from the output of the preamplifier to the probe.

## 4. Results and Discussion

Figure 2 shows the tuning curves generated by the spectrometer (Bruker 'wobb' command as implemented in Topspin 1.3 on a Bruker AV spectrometer), which represent a plot versus frequency of the difference in voltage drop between an ideal 50 Ω load and the circuit load. The cable length has a marked influence on the appearance of the tuning curves. As in previous investigations [2], it is found that the tuning curves rarely assume an ideal symmetrical "dip" form. A steadily increasing lobe on one side can be seen in different situations (either towards increasing or decreasing frequency values). For example, as can be seen in Figure 2, at cable lengths of 2λ and 2.1λ, the baseline of the tuning curve increases to higher frequencies and forms a shoulder below the tuning frequency. By contrast, moving to 2.2λ, the baseline of the tuning curve decreases to higher frequencies and forms a shoulder above the tuning frequency. This behavior repeats at every λ/2, as expected.

The SNTO position determined from the noise measurements is based on finding the tuning condition which achieves a symmetric dip line shape. It is seen that the SNTO is always found on the side of the increasing lobe in the tuning curve. In Figure 2 one can also see the NMR spin noise signals for each cable length at the CTO (conventional transmission tuning). It is found that the negative excursion of the signal occurs at higher frequencies when the SNTO is found at lower frequencies, and vice versa (note that by usual convention frequency decreases from left to right in the spectra, but increases from left to right in the tuning curves).

Between cable lengths of 2.1λ and 2.2λ, a transition occurs, rather abruptly, and a symmetric tuning curve can be obtained (Figure 2f). At this point, the SNTO position cannot be determined from noise measurements. In previous studies [7], such a situation has also been observed with several probe/preamplifier combinations. In Figure 2f, a tuning curve using the same cable length is shown. The tuning curve is a symmetrical curve in this regime. In addition, the spin-noise signal at CTO appears to be close to a perfect bump. This effect can be explained intuitively below in combination with insights about radiation damping factors.

Figure 3 shows the dependence of the SNTO position on the cable length used. It is seen that by changing the cable length, one may reach a regime in which both SNTO and CTO coincide (in this case at 2.3λ and 2.8λ), as also suggested in Reference [1]. Abrupt changes are seen at approximately 2.1λ and 2.6λ, where the SNTO offset changes from a large positive to a large negative offset. Simulations based on Eqs. (5)-(9) were then performed in MATLAB using the parameters $Q = 400$ (a typical measured value, see below), $L = 40$ nH, and $\omega_0 = 2\pi 500$ MHz. The circuit resistance $r$ was obtained via $r = \frac{\omega_0 L}{Q}$ as 0.314 Ω. In addition, $r_{s0}$ (Eq. (6)), the parameter quantifying the spin-noise resistance in relationship to $r$, was determined as 0.266 Ω from Figure 2f(ii), where a preamplifier noise figure of 1.1 was assumed. Using a spectrum analyzer to determine the preamplifier impedance $Z_p$ gave values in the range of 70 to 80 Ω with approximately 30 Ω reactive contributions (measured at 20 dBm). These measurements are likely incorrect because the power used by the analyzer probably saturated the preamplifier. Also, values of 500 Ω are much more common for NMR spectrometer preamplifiers and the simulations produced a much better fit with this value. One more adjustment needed to be made to fit the simulated curve well to the experimental data: an additional transmission line length of 0.4λ had to be added. It is easy to rationalize that such an additional line length could account for the internal electrical line lengths of the probe and the preamplifier module.

The results of the simulation are shown in the solid blue line in Figure 3 and are demonstrated to fit well the experimentally matched trends. Some ripples are seen in the simulated curve, likely as a result of instabilities in the minimization algorithm used for finding the SNTO condition.

Figure 4a shows the experimentally determined 'quality factor' $Q'$ as calculated by dividing the resonance frequency by the width of the tuning curve at half height (the tuning curve is represented in terms of voltage on the spectrometer). This parameter is measured as a function of transmission line length (when tuning to the CTO frequency). The maxima of this curve show the configurations at which radiation damping effects are strongest. The simulated curve is represented as a solid blue line and is based on calculating $1/\tau$ from Eq. (17). The vertical scaling of this simulated curve is treated as an adjustable parameter due to the many empirical constants that enter this equation. The shape of the curve, however, clearly follows the experimental curve, displaying both maxima and minima of $Q'$. This effect also illustrates that it is essentially unreliable to assess $Q$ factors via reflection coefficient measurements.

The symmetric spin noise line shape of Figure 2f(ii) was obtained in a regime, where radiation damping is lowest, thereby indicating that the overall resistance in the network is maximal. One can then explain the appearance of a 'bump' spin noise line shape as follows: From the Nyquist relation one obtains the voltage spectral density, which includes both the circuit and the spin contribution to the resistance. In order to obtain current, however, one divides by the absolute square of the total inductance, in which the contribution of the spin-noise becomes minimal. Hence the numerator in this expression becomes dominant (consisting of the sum of $N_r$ and $N_s$) and leads to the appearance of a 'bump' signal.

As outlined above, and described previously [2, 21, 22], strong radiation damping should lead to large frequency shifts of the signals. In our case, we have seen shifts of up to ~20 Hz. Figure 4b shows the frequency shifts of the resonance lines taken at all the sampled cable length positions, and shows a comparison with the frequency shift curve as calculated from Eq. (18). As with $Q'$, here the vertical scaling of the simulated curve was taken as an adjustable scaling factor, since a number of experimental parameters enter the equation, which are difficult to determine independently. A good correlation with the trend in $Q'$ values is found, thus illustrating the link between $Q'$ and radiation-damping-induced shifts. Notably, the largest shifts are found where $Q'$ is maximal, but in this region, they also switch from large positive to large negative shifts. Zero shifts can therefore also be found in this region by carefully adjusting the cable length. Alternatively, zero shifts can also be found at the minima of $Q'$, as one would expect. The variability of the shifts is larger at maximum $Q'$. At $Q'$ maxima, it is seen by comparison with Figure 3 that the SNTO can be made to coincide with the CTO, while at $Q'$ minima, the SNTO cannot be determined as it switches from a large positive to a large negative offset.

For a given line length, the frequency shift can be changed by tuning and matching off-resonance. Figure 4c shows the range of frequency shifts that can be observed in this way at a line length of $2.4\lambda$, where $Q'$ is approximately maximal. The experimental points are also compared with the simulated curve based on Eq. (19) and a good agreement can be found. This behavior is similar to what was observed in earlier studies [8, 22].

The signal-to-noise ratio (SNR) was assessed by performing one-dimensional pulse spectra with water suppression by selective presaturation using a 2 mM sucrose sample dissolved in 9/1 $H_2O/D_2O$. The SNR values were recorded for cable lengths of $2\lambda$ to $2.5\lambda$ (one half-cycle) at both the SNTO and CTO settings (Figure 5). It was found that the SNR fluctuated somewhat for the CTO setting (Figure 5b), but did not show pronounced maxima or minima. This behavior is indeed expected if the circuit is tuned and matched very close to 50 Ω. The fluctuations there indicate that the circuit cannot be matched exactly to 50 Ω, and/or that there is additional uncertainty in the SNR assessment, likely as a result of the changes in the water suppression conditions when the cable length is varied. At SNTO, the pulses were recalibrated for each cable length and water suppression conditions were optimized in each case. It was found here that the largest SNR appeared for cases where the SNTO was close to the CTO. The maximum SNR in this case also was similar to the SNR determined under CTO tuning, which supports the assumption that the impedance for optimal SNR at point B, $Z_{opt}$, is very close to 50 Ω. The solid blue line in Figure 5a shows the simulated SNR curve according to Eq. (11) with the vertical scale being an adjustable parameter. Since the experimental curves were acquired after saturation of the water resonance, $z_s$ was set to zero for the SNR calculation in these simulations after the SNTO tuning condition was found. In addition, it was found that the simulated SNR curve was offset by $-0.2\lambda$ from the maxima and minima of the experimental curve. The origin of this difference in cable lengths is not known at present. Changes in $Z_{opt}$ of both the real and imaginary parts would not explain such a shift as can be verified from Eq. (11). The appearance of one maximum and one minimum in the SNR curve per half wavelength is well understood based on this treatment.

The maximum SNR is found approximately where $Q'$ is maximal, and where hence the frequency shifts are minimal. Given the very different derivations of SNR and $Q'$, however, one can say that this finding is rather coincidental. The position of optimal SNR may further be affected by a change in $Z_{opt}$ via the noise sources $I_n$ and $V_n$, as has been pointed out previously [18, 19]. It would be desirable for many applications to find settings where the SNR would be maximal and the radiation damping minimal (minimal $Q'$). Such modifications could in theory be performed irrespective of the preamplifier input impedance $Z_p$, but in practice, there are limits to the range of such adjustments. The use of additional impedance transformation circuits within, before, or after the transmission line, which would operate asymmetrically in the forward and reverse direction [1] could offer additional flexibility.

## 5. Conclusions

We have explored the influence of the length of the coaxial cable that connects the preamplifier and the probe on a number of parameters, such as experimentally observable quality factors, sensitivity, radiation-damping-induced frequency shifts, and the noise (absorbed circuit noise and spin-noise) line shapes. It is found that by changing the cable length one can find a tuning regime where the maximum SNR is achieved, while the optimal transmission setting has a different optimum. By changing the cable length, it is shown that one can also make the optimal transmission and optimal noise reception settings coincide, which however leads to a slight reduction in SNR. The spin-noise and absorbed circuit noise spectral line shapes have a marked dependence on the tuning curves, and at certain cable lengths, the perfect SNTO line shape cannot be found. Radiation-damping-induced frequency shifts are seen to correlate with the cable lengths in a similar fashion as the experimentally observed quality factor values do. The SNTO is not always located where the frequency shift is minimal, nor is it always indicative of SNR optima. Most of these effects can be explained by the coaxial cable acting as a two-way impedance transformer, which matches the impedances at both the probe circuit, as well as, the preamplifier. The practical considerations described herein may prove useful for the optimization of different spectrometer setups for radiation damping blocking, transmission, or reception, or all combined. On commercial cryogenically cooled probes, however there are fewer options available for optimization since the cold part of the preamplifier is rigidly connected to the probe.

**Acknowledgments**

This work was supported by the US NSF grant to A.J (CHE-0957586), Austrian Science Funds FWF, project P 19635-N17 to N.M., and by the European Union (FP7 EAST-NMR, Contract No. 228461). The authors also would like to acknowledge discussions with Gareth Morris regarding a device that could perform an optimal impedance transformation in both the forward and backward directions between the probe and the preamplifier, and with Giovanni Boero on preamplifier noise sources.

**Figure Captions**

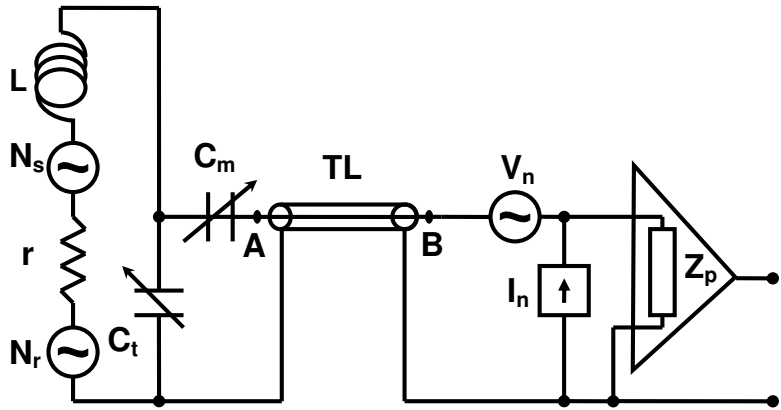

**Fig. 1.** Electronic model of the receiving coil and a preamplifier connected via a transmission line (TL). From the receiving coil there are two noise voltage sources, one associated with the resistance $r$ of the coil ($N_r$) and the other associated with the spin noise from the sample ($N_s$). $C_t$ and $C_m$ are the tuning and matching capacitors, respectively. The preamplifier has both voltage and current noise sources $V_n$ and $I_n$, respectively, where $Z_p$ is the preamplifier input impedance.

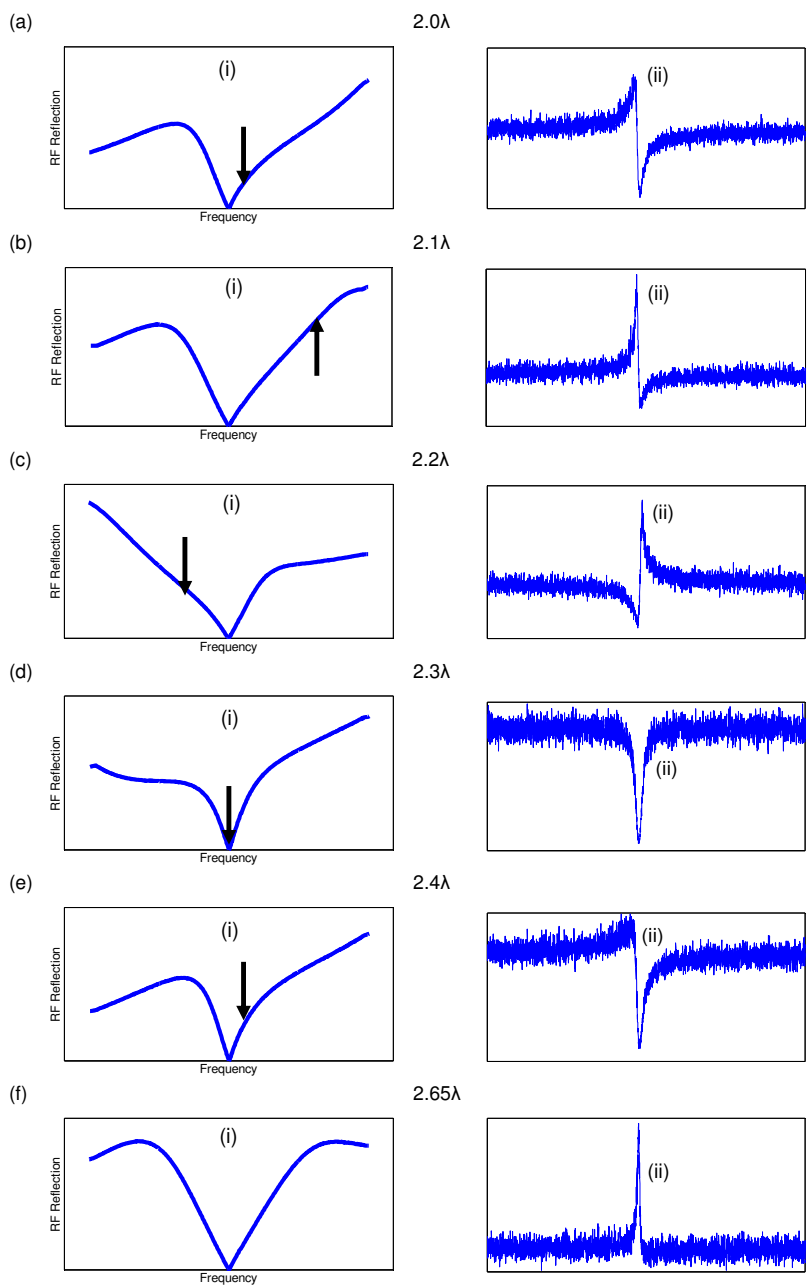

**Fig. 2.** Comparison of tuning curves and spin-noise signals at 500 MHz using the indicated coaxial cable lengths between the preamplifier to the probe. (i) Tuning curves tuned and matched to the CTO at 500.202 MHz. The arrows indicate the SNTO position (ii) Corresponding spin-noise signals at CTO. **(a)** 2.0λ cable length, SNTO at 500.907 MHz (+705 kHz shift from CTO) **(b)** 2.1λ cable length, SNTO at 502.400 MHz (+2198 kHz shift from CTO) **(c)** 2.2λ cable length, SNTO at 498.837 MHz (-1365 kHz shift from CTO) **(d).** 2.3λ cable length, SNTO at 500.202 MHz (0 kHz shift from CTO) **(e).** 2.4λ cable length, SNTO at 500.594 MHz (+392 kHz shift from CTO) **(f)** 2.65λ cable length, no SNTO found.

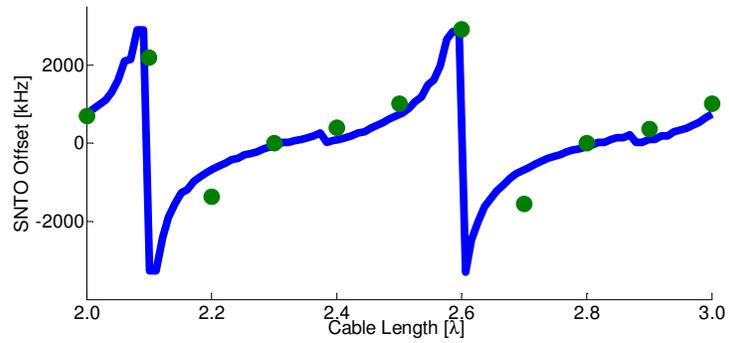

**Fig. 3.** SNTO offset from CTO as a function of coaxial cable length in units of wavelength. Green circles: Measured values, where the SNTO positions were found by determining the tuning frequency that gave a symmetrical dip of the water proton spin-noise signal. Blue line: Simulated curve, values of $L$=40 nH, $Q$=400 were used in the simulation. Additional simulation parameters are described in the text.

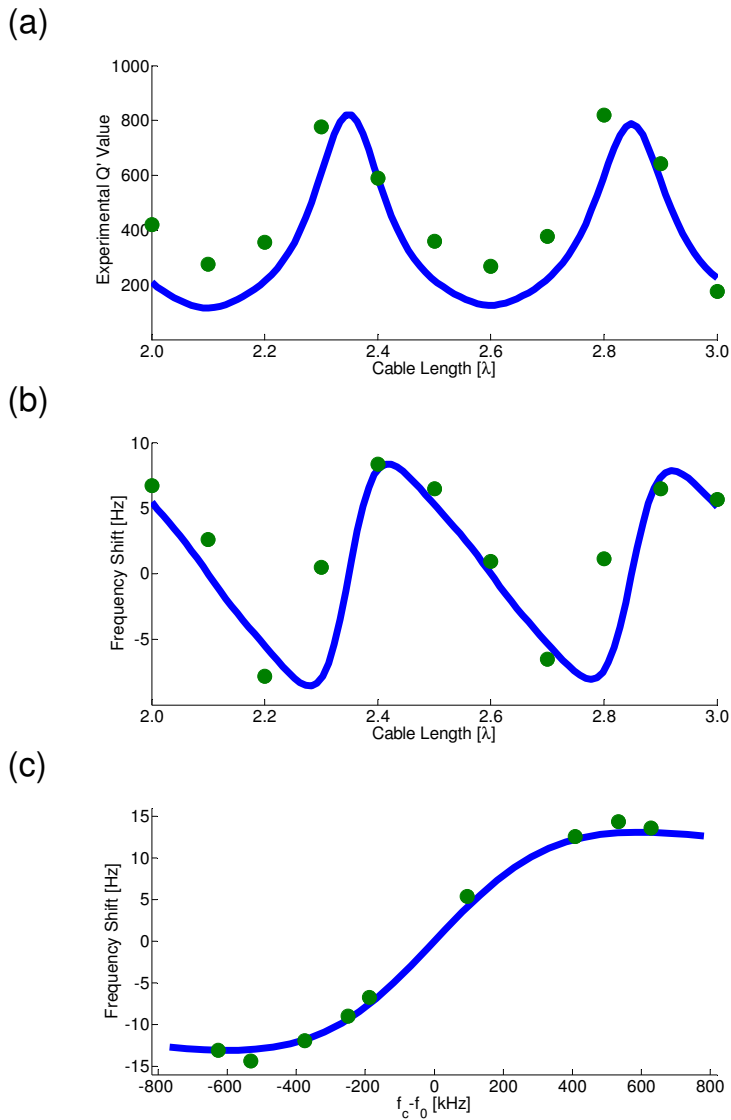

**Fig. 4. (a)** Experimental $Q'$ values as a function of cable length in units of wavelength. Green circles: Measured $Q'$ values. $Q'$ was calculated using the tuning frequency (all were done at CTO frequency), divided by the width of the tuning curve at half height from the baseline. Blue line: Simulated $Q'$ values **(b)** Resonance line frequency shift at CTO as a function of coaxial cable length in units of wavelength. Green circles: Measured values. Blue line: Simulation. **(c)** Frequency shifts of water proton signals using a small angle RF pulse of ($5^0$). A 2.4λ coax cable length was used. Green circles: Measured values. Blue line: Simulation based on Eq. (19) The frequency shift is plotted as a function of tuning offset $f_c-f_0$ for values in the range of $-800$ kHz $< f_c-f_0 < 800$ kHz

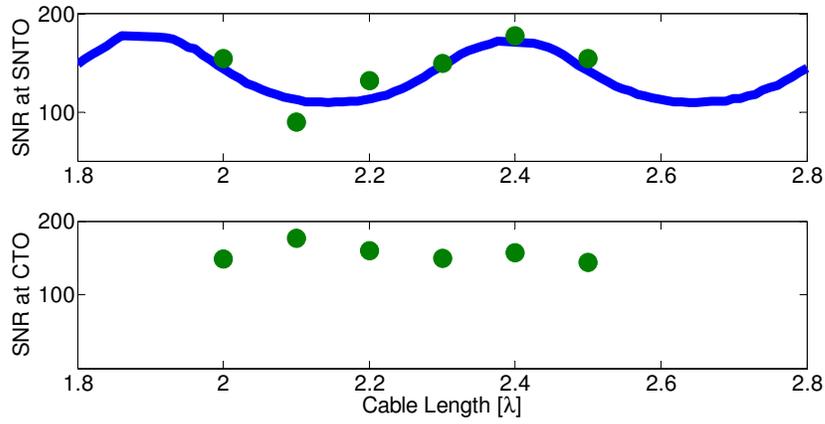

**Fig. 5.** SNR values as a function of cable length in units of wavelength. Green circles: Measured SNR values at (a) SNTO, and (b) CTO. Blue line: Simulated SNR values (for SNTO tuning plot in (a) only). The simulated plot is shifted by -0.2λ to obtain the best fit as described in the text.